\documentclass[twocolumn,floats,amsmath,amssymb,prb]{revtex4}

\usepackage{graphicx}

\begin{document}

\title{Topological characterization of crystalline ice structures
       from coordination sequences}
\author{Carlos P. Herrero}
\author{Rafael Ram\'irez}
\affiliation{Instituto de Ciencia de Materiales de Madrid,
         Consejo Superior de Investigaciones Cient\'ificas (CSIC),
         Campus de Cantoblanco, 28049 Madrid, Spain }
\date{\today}

\begin{abstract}
Topological properties of crystalline ice structures are studied by
considering ring statistics, coordination sequences, and topological
density of different ice phases.  The coordination sequences 
(number of sites at topological distance $k$ from a reference site) 
have been obtained by direct enumeration until at least 
40 coordination spheres for different ice polymorphs.
This allows us to study the asymptotic behavior of the mean number of 
sites in the $k$-th shell, $M_k$, for high values of $k$:
$M_k \sim a k^2$, $a$ being a structure-dependent parameter. 
Small departures from a strict parabolic dependence have been 
studied by considering first and second differences of the series 
$\{M_k\}$ for each structure.
The parameter $a$ ranges from 2.00 for ice VI to 4.27 for
ice XII, and is used to define a 
topological density for these solid phases of water.
Correlations between such topological density and the actual volume of 
ice phases are discussed. Ices Ih and Ic are found to depart from the
general trend in this correlation due to the large void space in their
structures.
\end{abstract}

\maketitle

\section{Introduction}

Water presents a large variety of solid structures, and
up to now sixteen different crystalline ice phases have been
found.\cite{du10,ba12,sa11,zh06}
A large amount of experimental and theoretical work has been
devoted to determine precisely their crystal structures and
stability range in the pressure-temperature phase diagram.
Notwithstanding the broad knowledge so obtained, some properties 
of these solid phases still lack a full understanding. 
This is mainly owing to the presence of hydrogen bonds between 
contiguous molecules, which gives rise to some peculiarities  in
their properties (the so-called `water anomalies').\cite{ei69,pe99,ro96}

Water molecules appear in all known ice phases (with the exception of
ice X) as well defined entities building up a network linked by H-bonds. 
In such a network each water molecule is bound to four others in a more 
or less distorted tetrahedral coordination. 
The orientation of each molecule with respect to its four nearest 
neighbors complies with the so-called Bernal-Fowler ice rules. 
These rules state that each H$_{2}$O molecule is oriented
so that its two hydrogen atoms point toward contiguous oxygen atoms
and there is exactly one hydrogen between two adjacent oxygen
atoms.\cite{be33} 

Orientational disorder of the water molecules appears in
several ice phases. Though oxygen atoms display
full occupancy ($f$) of their crystallographic sites,
hydrogen atoms may present a disordered distribution, as
shown by a fractional occupancy of their lattice positions.
Thus, hexagonal ice Ih, the stable phase of solid water under normal 
conditions, presents full hydrogen disorder compatible with the ice rules
(occupancy of H-sites $f = 0.5$).
However, some phases such as ice II are H-ordered, whilst others as
ices III and V are characterized by partial hydrogen order
(some fractional occupancies $f \neq 0.5$).

Given the variety of ice structures, some unifying classification
can help to deeper understanding of their physical and chemical 
properties.\cite{ma09,sa11}
Most classification schemes of crystalline solids are based on the
space symmetry and/or the short-distance atomic environments.
In other schemes, attention is focused on geometrical aspects of
packing of structural units.
Such classification methods can be considered as geometrical, as their
main criteria are geometrical properties of crystal
structures.\cite{pe72,we86,li85,bl00}
A difficult issue of these geometrical classification methods can be
the involvement of finding relations between compounds whose structures
are distorted.
Another possibility is the use of classification schemes based on
topological criteria, i.e., putting emphasis on the organization of the
interatomic bonds in a crystal structure as a basic criterion for a
crystal-chemical analysis.
Thus, topological properties of crystal structures have been
taken into account in the past to describe different kinds of
materials.\cite{pe72,li85,we86}

A discussion of the network topologies of different ice polymorphs
and the relation of ring sizes in the various phases with the crystal
volume has been presented by Salzmann {\em et al.}\cite{sa11}
Topological studies of three-dimensional hydrogen-bonded frameworks in
organic crystals have also helped to classify this kind of
structures.\cite{ba07}     Moreover,
the topology of hydrogen-bond networks in ice has been considered
in order to analyze hydrogen ordering.
In particular, graph invariants have turned out to be very useful to
obtain the energy of hydrogen configurations on a given ice
network.\cite{si12,kn06,kn08}
Graph theory has been also used to study isotypism in crystal
structures.\cite{bl00}

To characterize the ice structures from a topological point of view,
we will employ the so-called coordination sequences.
This is a generalization of the coordination number, usually known as
the number of nearest neighbors of an atom in a solid structure.
In this respect, an ice structure can be viewed as a three-dimensional
hydrogen-bonded network, and thus one can consider a simplified structure 
where each oxygen atom is assumed to be linked to four other oxygen atoms,
without explicit mention of the hydrogen atoms lying between them.
Then, for a particular site,
one defines a coordination sequence as a series of numbers $\{N_k\}$
($k$ = 1, 2, ...), where $N_k$ is the number of sites located at a
topological distance $k$ from the reference (see below).
Note that this is a purely topological concept, as the coordination
sequence for each oxygen atom in an ice structure depends only on the
topology of the network, but not on lattice distortions and other
structural factors.

   The concept of coordination sequence was applied to silicate
frameworks by Meier and Moeck in 1979.\cite{me79}
Since then, several authors employed this concept to
characterize from a topological point of view different types of
materials\cite{br79,co97,gr96,st90,he94,eo02}
and complex networks.\cite{he02b}
Coordination sequences can be used to define a topological density, 
as a structural characteristic related to the
increase in the number $N_k$ of sites accessible through $k$ links in
a given structure.
In this paper, we analyze this question by calculating the coordination
sequence for crystalline phases of ice, up to at least $k$ = 40,
and thus approaching the asymptotic behavior of $\{N_k\}$ at 
large distances. A definition of topological density can be derived 
from such an asymptotic behavior.   

This kind of topological studies are relevant to connect the small-scale
characteristics of solid water to macroscopic features such as mechanical
or plastic properties, not only for crystalline structures but also
for various types of amorphous ice, as well as liquid water.\cite{pe99}
Moreover, networks of water molecules are paradigmatic in the study
of compounds for which H-bonds play a significant role in their
structural, dynamical, and electronic properties.

\section{Computational method}

Our model is defined in the following way. We consider an ice structure 
as defined by the positions of the oxygen atoms, so that each O atom has 
four nearest neighbors. This defines a network, where the nodes are the O
sites, and the links are H-bonds between nearest neighbors.
The network coordination is four, which gives a total of $2N$ links,
$N$ being the number of nodes. We implicitly assume that on each link 
there is one H atom, but its consideration is not relevant for our
present purposes.

For a given network,
we define the `topological distance' between two sites as the number of
bonds in the shortest path connecting one site to the other.
This is obviously a symmetric property, in the sense that the 
topological distance from site A to site B is the same as that from
B to A. We will call $N_k$ the number of sites at a topological
distance $k$ from a given node, i.e. in its $k$'th `coordination
shell'.
This is a generalization of the usual coordination shell formed by 
nearest neighbors in solid structures ($k$ = 1). 
Then, we call the sequence $\{N_k\}$ the `coordination sequence' 
of this node. 
In general, different nodes in a network may have different 
coordination sequences, and we will call coordination sequence of a 
network to the sequence $\{M_k\}$, where each $M_k$ is obtained
by averaging the $N_k$ values for the oxygen sites in the unit cell.
For ice structures including O sites topologically non-equivalent
(e.g., ices III, IV, V, VI, and XII), 
relative differences between $N_k$ values corresponding to
sites in a given structure decrease fast with the distance $k$. In fact,
the relative difference is about 0.1\% for $k$ = 40, and becomes
negligible in the large-$k$ limit.

%  TABLE 1
\begin{table}[ht]
\caption{Crystal system and space group for the ice polymorphs
considered in this work.\\}
\centering
\setlength{\tabcolsep}{10pt}
\begin{tabular}{cccc}
  \hline 
  \vspace*{0.1cm}
   Phase & Crystal system & Space group &  Ref. \\
   \hline \vspace*{-0.2cm} \\
   Ih    & Hexagonal    & $P6_3$/$mmc$, 194 &  \onlinecite{pe57} \\
   Ic    &   Cubic      & $Fd\bar{3}m$, 227 &  \onlinecite{ko44} \\
   II    & Rhombohedral & $R\bar{3}$, 148   &  \onlinecite{ka71} \\
   III   & Tetragonal   & $P4_12_12$, 92    &  \onlinecite{lo00} \\
   IV    & Rhombohedral & $R\bar{3}c$, 167  &  \onlinecite{en81} \\
   V     & Monoclinic   & $A2/a$, 15        &  \onlinecite{lo00} \\
   VI    & Tetragonal   & $P4_2/nmc$, 137   &  \onlinecite{ku84} \\
   XII   & Tetragonal   & $I\bar{4}2d$, 122 &  \onlinecite{lo98} \\
   &&& \vspace*{-0.3cm} \\
   \hline
\end{tabular}
\label{tb:ice_crystall_systems}
\end{table}

For some solid structures, it is known that one can find regularities
in the sequence $\{M_k\}$ for relatively low values of the
topological distance $k$,\cite{gr96}  and thus easily extrapolate to
the behavior of $\{M_k\}$ at large distances.
In general, however, a study of the asymptotic behavior of coordination 
sequences for a given network requires the generation of large supercells.
    For the ice structures considered in this work, supercells
including around $3 \times 10^5$ oxygen sites
were generated from structural data taken from the literature. 
The relevant information on these structural data is presented in Table~I. 
Then, given an oxygen site, numbers of sites $N_k$ in successive
coordination shells were obtained by direct enumeration.  
To this end, one only needs to assign a label (e.g., a number) to each 
node in the considered supercell, and associate it with those 
corresponding to its four nearest neighbors. 
Thus, given a node, one finds successively
its first neighbors, then the neighbors of these, and so on, with the
precaution of not going `backwards', i.e., not counting again sites
already visited in a previous step.

It is important to realize that the coordination sequence 
for each site in a given structure depends only on the topology of
the network, and not on the actual symmetry, cell parameters, or
other structural data. In particular, the coordination sequence $\{M_k\}$
for a given ice structure is not affected by the ordering of H atoms.
This means that structures such as those of ices Ih and XI (the former
being H-disordered and the latter H-ordered), with the same topology,
will have the same sequence $\{M_k\}$.
There are five other pairs of ice structures sharing the same topology,
and connected one with the other through an order/disorder phase
transition.  Thus, one has six pairs of ice structures: 
Ih-XI, III-IX, V-XIII, VI-XV, VII-VIII, and XII-XIV.\cite{sa11} 
In each pair, the first structure is H-disordered, and
the second is H-ordered. 
In the following, for our topological characterization we will only
refer to the disordered case, but meaning that it represents both
members of the corresponding pair.
Apart from these, there are ice structures with topology different from
that of the structures above, and for which no pair has been found:
ice Ic (H-disordered), ice II (H-ordered), and ice IV (H-disordered).  
There is also the structure of ice X, which is topologically equivalent 
to those of ices VII and VIII, but with the main difference that in ice X
hydrogen atoms lie at the middle point between oxygen atoms, and 
water molecules lose in fact their own entity (this is however irrelevant
for our topological characterization).
Then, we have 9 topologically different ice structures. 
We note that the network associated to ice VII 
(as well as ices VIII and X) is made up of two
interpenetrating but disjoint subnetworks, each of them
equivalent to the ice Ic network, and therefore with its same topology.
There is another case, the pair VI-XV, for which the network consists of
two independent subnetworks, but they are not equivalent to the network 
of any other known ice structure.
For these reasons we will consider the 8 structures listed in 
Table~I, where we indicate the crystal system, space group, and
the reference we used to generate the corresponding supercells.

Coordination sequences corresponding to structures in $D$-dimensional
space increase with topological distance $k$ roughly 
as $k^{D-1}$.\cite{co97,br79} 
Then, for three-dimensional structures one expects for large 
$k$ a dependence
\begin{equation}
         M_k \sim a \, k^2     \hspace{3mm}  ,
\label{mk}
\end{equation}
where $a$ is a network-dependent parameter. Thus, $M_k$ increases 
quadratically with $k$ just as the surface of a sphere increases
quadratically with its radius.
This is in fact the dependence found for different kinds of real
materials.\cite{he94,gr96}
In general, including also small values of $k$, a quadratic dependence
of the form
\begin{equation}
     M_k = a k^2 + b k + c    
\label{nk}
\end{equation}
has been found to follow closely the coordination sequences of 
actual structures.\cite{he94,gr96}
Given a dependence such as Eq.~(\ref{nk}), the parameters $b$ and $c$
become irrelevant for large distances, and one has 
$M_k/k^2 \to a$.
Such a parabolic dependence, although it follows closely the sequence
$\{M_k\}$ for solid structures, is not strictly followed in general.
Thus, it is well known for the diamond structure (ice Ic in our case)
that the equation
$M_k = 2.5 \, k^2 + 1.75$ yields values too high by 0.25 if $k$ is odd
and too low by 0.25 if $k$ is even.\cite{br79,gr96}

To characterize the coordination sequences of ice structures, and
in particular their deviation from strict parabolic behavior,
we will consider first and second differences.
We define the first differences of a sequence $\{M_k\}$ as
\begin{equation}
  \delta_k = M_{k+1} - M_k
\label{dif1}
\end{equation}
and the second differences as
\begin{equation}
   \epsilon_k =  \frac12 ( \delta_k - \delta_{k-1} ) 
              = \frac12 (M_{k+1} - 2 M_k + M_{k-1})
\label{dif2}
\end{equation}
Note the factor $1/2$ in Eq.~(\ref{dif2}), introduced for the sake of 
comparison with the parameter $a$ in Eq.~(\ref{nk}). 
In fact, for a strict parabolic dependence of $M_k$ on $k$, as in 
Eq.~(\ref{nk}), one would have a constant second difference  
$\epsilon_k = a$ for all $k$.

Since the parameter $a$ is a quantitative measure of the mean number of 
sites connected to a given site in a finite number of steps on the 
corresponding network, we define the `topological density' $\rho$ as
\begin{equation}
   \rho = w \, a  \,  ,
\label{rho}
\end{equation}
where $w$ is the number of disconnected subnetworks in the
considered network. Usually $w = 1$, but for the ice structures
including two interpenetrating networks (as ices VI and VII) one 
has $w = 2$.

For a given network, 
we call $S_k$ the mean number of sites up to coordination shell $k$,
without counting the starting site (this has been called in the
literature\cite{co97} a `crystal ball' of radius $k$):
\begin{equation}
    S_k = \sum_{i=1}^k M_i
\label{sk}
\end{equation}
This quantity has been employed earlier to quantify the topological
density of crystal structures.\cite{br93}

To investigate the relation between topological density and 
real density of ice structures, we will compare values of 
$\rho$ with those of the volume per molecule $v$.
In this respect, it is important to remember that the different ice
structures are stable in some cases at very different conditions of
pressure and temperature. This means that a direct comparison of 
$v$ values for different phases with their corresponding topological
densities may be misleading, as $\rho$ is independent of bond lengths,
lattice distortions, and other structural characteristics that may
dramatically change with pressure and/or temperature.

To avoid this problem, we will consider for each ice structure 
a reference volume $v_0$, defined as the volume per molecule that 
minimizes the potential energy of the crystal at zero pressure.
For this purpose we need a reliable potential model.
We have used the point charge, flexible q-TIP4P/F model, developed 
to study liquid water,\cite{ha09} and that has been used later to study 
various properties of ice\cite{ra10,he11} and water clusters.\cite{go10}
This interatomic potential takes into account the significant
anharmonicity of the O--H vibration in a water molecule by considering
anharmonic stretches.
The q-TIP4P/F interaction model has turned out to give reliable
results for several ice phases, and in particular it predicts crystal 
volumes in fairly good agreement with experimental data.\cite{he11,ra12}
For each ice structure, determination of the reference volume was
carried out by an energy minimization, where both atomic positions 
and cell parameters were optimized. 
Details on this minimization procedure can be found 
elsewhere.\cite{ra12}

%  TABLE 2
\begin{table*}[ht]
\caption{Number and size ($L$) of the rings appearing in different
ice structures.
$m_j$ is the label of the crystallographic position of the oxygen
atoms. The numeral in the label indicates the multiplicity of the
site in the considered unit cell. For ices II and IV the given
crystallographic positions correspond to the choice of a
rhombohedral cell in the space group.
$x(L)$ indicates the fraction of loops in an ice structure for
each size $L$. $\langle L \rangle $ is the mean loop size, as
defined in Eq.~(\ref{meanl}).\\}
\centering
\setlength{\tabcolsep}{8pt}
\begin{tabular}{cccccccccccc}
  \hline \hline \vspace*{-0.2cm} \\
  \vspace*{0.1cm}
Ice &  Site $j$ & $m_j$ & & & &  $L$  &  &  &  &  &
          $\langle L \rangle $  \\
\cline{4-11}
  &  &  &   4  &  5 &   6 &   7 &   8 &   9 &  10 &   12   \\
   \hline \vspace*{-0.2cm} \\
Ih  &  1 &   4f &    - &   - &  12 &   - &   -  &  - &   - &    - \\
    &  2 &   4f  &   - &   - &  12  &  -  &  -  &  -  &  - &    - \\
& $x(L)$  & & & &       1.0  &  & & & & &  6 \\
   \hline \vspace*{-0.2cm} \\
Ic  & 1  &  8a   &  -  &  - &  12 &   - &   - &   - &   - &    - \\
& $x(L)$ & & & &       1.0 & &  & & & &  6 \\
   \hline \vspace*{-0.2cm} \\
II  & 1 &   6f &    - &   - &   7 &   - &  12 &   - &  25 &    - \\
    & 2 &   6f &    - &   - &   7 &   - &  12 &   - &  25 &    - \\
& $x(L)$  & & & &     0.226 &  &   0.290 & &   0.484  & &   8.52 \\
   \hline \vspace*{-0.2cm} \\
III  & 1 &   4a &    - &   4 &   - &   4 &   4 &   - &   - &    - \\
     & 2 &   8b &    - &   3 &   - &   5 &   6 &   - &   - &    - \\
& $x(L)$ & & &   0.333 & &   0.333 & 0.333  & & & &   6.67  \\
   \hline \vspace*{-0.2cm} \\
IV  & 1 &   4c &    - &   - &   6 &   - &  12 &   - &  63 &    - \\
    & 2 &  12f &    - &   - &   5 &   - &  20 &   - &  49 &    - \\
& $x(L)$  & & & &     0.104 & &   0.269  & &   0.627   & &  9.04  \\
   \hline \vspace*{-0.2cm} \\
V  & 1 &   4e &    - &   3 &   2 &   - &   2 &   4 &  24 &    2 \\
   & 2 &   8f &    1 &   3 &   2 &   - &   3 &   2 &  16 &    1 \\
   & 3 &   8f &    1 &   1 &   2 &   - &   4 &   3 &  18 &    2 \\
   & 4 &   8f &    2 &   2 &   1 &   - &   4 &   2 &  14 &    2 \\
& $x(L)$ & & 0.08& 0.12& 0.08&    &  0.12& 0.08& 0.48 &   0.04 &
8.36  \\
   \hline \vspace*{-0.2cm} \\
VI  & 1 &   2a &    4 &   - &   - &   - &  16 &   - &   - &    - \\
    & 2 &   8g &    4 &   - &   - &   - &  14 &   - &   - &    - \\
& $x(L)$ & &  0.357  &  &  &  &   0.643  & & & &   6.57  \\
   \hline \vspace*{-0.2cm} \\
XII  & 1 &   4a &    - &   - &   - &   8 &  16 &   - &   - &    - \\
     & 2 &   8d &    - &   - &   - &  10 &  16 &   - &   - &    - \\
& $x(L)$  & & & & &          0.4 &  0.6  & & & &   7.6   \\

   &&&& \vspace*{-0.3cm} \\
   \hline \hline
\end{tabular}
\label{tb:loops}
\end{table*}

\section{Results and discussion}

\subsection{Rings}

Given the variety of known crystalline ice structures, it is not strange
that the coordination of water molecules further than nearest neighbors
may display clear differences between different structures, as could be
expected from the pressure/temperature range where they are stable.
In this line, as a first characteristic observable in the ice networks, 
we consider rings of water molecules, which define the particular
topology of each structure.
A discussion of the structural variation in tetrahedral networks
was presented by Zachariasen as early as 1932,\cite{za32} and a systematic
description of the topology of different network types was given by
Bernal,\cite{be64} who suggested that rings should be a basic topological
measure, in particular for covalently bonded
structures.\cite{ki67,st90b,gu90}

In general, a ring may be defined as any returning path in a network.
This definition is however not very useful, as there is an infinite number 
of those rings. For practical reasons, it is desirable to identify a
small set of rings as fundamental topological units.
Following Marians and Hobbs,\cite{ma90} we will define a fundamental ring 
as any ring that cannot be divided into two smaller ones.
Thus, in the computer code employed to analyze the ice structures,
we first find loops in a given network, and then we check for each one 
that there is no possible `shortcut' between any pair of sites in the 
loop, as this would mean that the loop could be divided into two 
smaller ones. This definition coincides with that employed by 
Salzmann {\em et al.}\cite{sa11} to analyze correlations between ring
size and density of ice polymorphs. 
Note that we call simply rings (or structural rings) what has been 
called `minimal rings' by Guttman,\cite{gu90} `strong rings' by 
Goetzke and Klein,\cite{go91} or `primitive rings' by Yuan and
Cormack\cite{yu02} in their studies of different types of materials
and topological networks.
Other definitions, non-equivalent to that given here, have been
presented in the literature. In particular, for Goetzke and
Klein,\cite{go91} a loop is a `very strong ring' if it contains some
link which is not a member of some shorter ring. This set of rings
is a subset of those considered here, and is interesting in graph
theory for its properties as a covering set, i.e., the set of very
strong rings of a graph is always a cover of all links of cycles of
the graph. This property and ring definitions related to it are of 
no particular interest for our present purposes, and thus 
our definition is taken to be consistent with those mentioned above 
and employed earlier to characterize several types of materials,
including ice polymorphs.\cite{sa11} 

Calling $n_j(L)$ the number of rings of size $L$ which include a given
site of type $j$, the number of $L$-membered rings per unit cell is
\begin{equation}
  r(L) = \sum_j m_j \, \frac{n_j(L)}{L}
\label{ri}
\end{equation}
where $m_j$ is the multiplicity of site $j$ in the unit cell. 
Then, the fraction $x(L)$ of rings of size $L$ is
\begin{equation}
  x(L) = \frac{r(L)}{R} \, , \hspace{1cm}   R = \sum_L r(L)
\label{xl}
\end{equation}
and the mean ring size $\langle L \rangle $ is given by
\begin{equation}
   \langle L \rangle  = \sum_L L \, x(L)
\label{meanl}
\end{equation}

For each structure and ring size, we present in Table~II the number of
rings which include a given network site. We give separately results for
the different crystallographic sites in each structure, as they may have
different topological environment.
It is clear that crystallographically equivalent sites are topologically
equivalent, and therefore have the same coordination sequence, but
sites crystallographically non-equivalent may be
topologically equivalent or not.
In several ice structures all sites are topologically equivalent
(homogeneous networks\cite{me79}), as we observe for ices Ih, Ic, and II,
where the number and size of the rings coincide for all sites in each
structure.     In other structures, however,
there appear several (up to 4) topologically distinct sites
(heterogeneous networks\cite{me79}),  and different crystallographic 
sites have different topological environments (see Table~II).

A distinct property of ices Ih and Ic is that all rings in their 
crystal structures have the same size (six-membered rings, $L = 6$).
Ices VI and XII display rings of two different sizes
($L$ = 4 and 8 for ice VI; 7 and 8 for ice XII).
The largest variety of ring sizes corresponds to ice V, for which
one finds rings of seven different sizes.
The minimum ring size found in the considered structures is four, 
which appears for ices V and VI.
The largest loops contain 12 sites, and only appear in ice V.
In Table~II we also present the mean ring size $\langle L \rangle $
for each ice
polymorph. The smallest mean size is six (for ices Ih and Ic), and the
largest amounts to 9.04 (for ice IV).
Average ring sizes smaller than six have been found in the water
networks of various clathrate hydrates.\cite{lo08}

It is known that the structure of water in confined regions
appreciably differs from those of the bulk liquid and solid phases,
and one may find peculiar topologies for the hydrogen-bond networks
in those cases.\cite{we11,zh10,by06,gh04} 
In particular, for water-filled carbon nanotubes
the H-bond network consists of stacked $L$-membered rings, where $L$ 
goes from 4 to 6 depending on the nanotube diameter.\cite{ko04b,ma05b}  
Infrared spectroscopy has allowed in this case to
detect distinct vibrational frequencies associated to intra- and 
inter-ring hydrogen bonds,\cite{by06} revealing this technique as a
reliable complementary tool to diffraction studies in the topological 
characterization of confined water.\cite{we11}
Similar procedures can be also used to study the actual topology of
various phases of amorphous ice, where experimental techniques and 
simulations have been shown to yield complementary information on this
matter.\cite{ma05b,ts05,sa06,fi02,tu02b,bo06}

%  TABLE 3
\begin{table*}[ht]
\caption{Average number of sites $M_k$  for
$k$ = 2, 3, 4, 10, 20, and 30, for different ice structures,
along with the corresponding parameter $a$ obtained from a fit
to the equation $M_k = a k^2 + b k + c$.  $S_{30}$ is the total
number of sites up to $k$ = 30.   $\rho / \rho_{\rm Ih}$ is the
relative topological density with respect to ice Ih.
On the last column, P and NP indicate whether the sequence
$\{\epsilon_k\}$ was found to be periodic or non-periodic; when a
period was found, a number indicates its length.
}
\vspace{0.3cm}
\centering
\setlength{\tabcolsep}{10pt}
\begin{tabular}{c c c c c c c c c c c}
 Ice  &  $M_2$  &  $M_3$ &  $M_4$ & $M_{10} $  & $M_{20}$ &
$M_{30}$
      &  $S_{30}$ & $a$  &  $\rho / \rho_{\rm Ih}$  &$ \{ \epsilon_k \}$ \\
\hline
 Ih            &  12  &  25  &  44   &   264  &  1052  &  2364
               & 24866  &  2.62  & 1 & P 4       \\ 

 Ic            &  12  &  24  &  42  &  252  &  1002  &  2252
               & 23690  &  2.50  &  0.95 &  P 2       \\ 

 II            &  12  &  29  &   58  & 346  &  1392  &  3136
               & 32897  &  3.50  &  1.34 &  P 4        \\ 

 III           &  12  &  28  &  49.3 & 319.3 & 1289.3 &  2907.3
               &  30474  &  3.24 &  1.24 &  NP          \\ 

 IV            &  12    & 31   &  65  & 402  &  1629.5  &  3681
               & 38486  &  4.12  &  1.57  &   NP        \\ 

 V             &  10.9  & 24.9 & 53.1 & 379.4 &  1530 &  3464.6
               & 36207.1  &  3.86 &  1.47  &   NP        \\ 

 VI            &  8.8   & 18.8 & 34.4 & 201.6 &  802.4 &  1800.4
               & 18956.8  &  2.00  &  1.53  &  P 12      \\ 

 XII           &  12   &  36  &  62  &  417.3 & 1686.7 &  3814.7
               &  39957.3 &  4.27  &  1.63  &  P 20       \\ 

\end {tabular}
\label{tb:coord_seq}
\end{table*}

\subsection{Coordination sequences}

We now turn to the coordination sequences of the different ice
structures.  In Table~III we give the mean values $M_k$ of the
coordination sequence of the considered structures for several
values of the topological distance $k$. We show in particular the
first terms in the sequence (for $k$ = 2, 3, and 4) to see the effect of
structural rings. We present also $M_k$ for larger $k$ values, such as
$k$ = 10, 20, and 30. The accumulated number of sites up to $k$ = 30
is given by $S_{30}$ for each structure.
The coordination sequence of ice Ih is the same as that of 
$\beta$-tridymite silica, as both structures have the same topology.
Also, the sequence $\{ M_k \}$ for ice Ic coincides with that of
$\beta$-cristobalite and diamond.\cite{br79}

\begin{figure}
\vspace{-0.7cm}
\hspace{-0.5cm}
\includegraphics[width= 9cm]{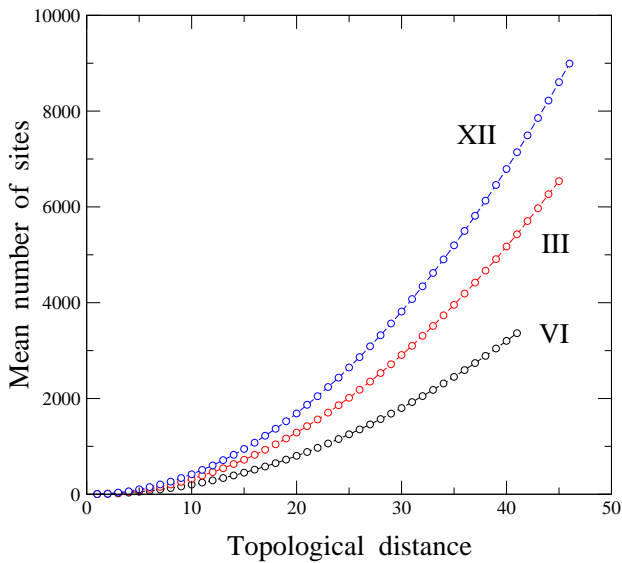}
\vspace{-0.7cm}
\caption{
Coordination sequences $\{ M_k \}$ of three ice structures
vs topological distance $k$.
From top to bottom: ice XII, III, and VI.
}
\label{f1}
\end{figure}

For a network without loops (Bethe lattice\cite{zi79,st90}), 
the coordination sequence is given by
\begin{equation}
    M_k^B = z (z - 1)^{k-1}
\label{bethe}
\end{equation}
where $z$ is the number of nearest neighbors (degree or connectivity
in the language of graph theory), assumed to be the same for all sites
($z = 4$ here).
For actual crystal structures, whose networks include loops, $M_k$ will be
in general smaller than $M_k^B$.
 The number of sites at topological distance $k$, $M_k$, is 
affected by the number of structural rings of size  $L \leq k/2$.
For example, the presence of a four-membered ring causes a reduction of one
node in the second coordination shell ($k = 2$); five- and six-membered 
rings affect the coordination sequence for $k \geq 3$, and so on.

The lower terms of the sequence $\{M_k\}$ are especially indicative 
of the relative abundance of small rings, as such rings appreciably 
contribute to decrease the $M_k$ values.
None of the known crystalline ice structures contain 3-membered rings. 
For structures not including 4-membered rings, one has
$M_2 = 12$ (the maximum value allowed by Eq.~(\ref{bethe}) ).
Ices V and VI contain 4-membered rings, and their
mean $M_2$ values are 10.9 and 8.8, respectively.
In the same way, 5- and 6-membered rings contribute to decrease 
$M_3$ from its possible maximum value ($M_3^B = 36$) for all considered
structures, with the exception of ice XII, for which the smallest loops
include seven water molecules (see Table~II).

In Fig.~1 we show the coordination sequence $\{M_k\}$ for three ice
structures (ices III, VI, and XII) up to $k \sim 40$. One observes the
apparent parabolic dependence for $\{M_k\}$ in all cases, but with
clearly different slopes. Thus, for ice XII, $M_{40}$ is more than twice
the corresponding value for ice VI, in agreement with the difference
obtained for the coefficient $a$ of $k^2$ in Eq.~(\ref{nk}) (4.27 for ice
XII vs 2.00 for ice VI; see Table~III).
For other ice structures, the coordination sequence has $M_k$ values 
intermediate between those corresponding to ices VI and XII. 

\begin{figure}
\vspace{-0.7cm}
\hspace{-0.5cm}
\includegraphics[width= 9cm]{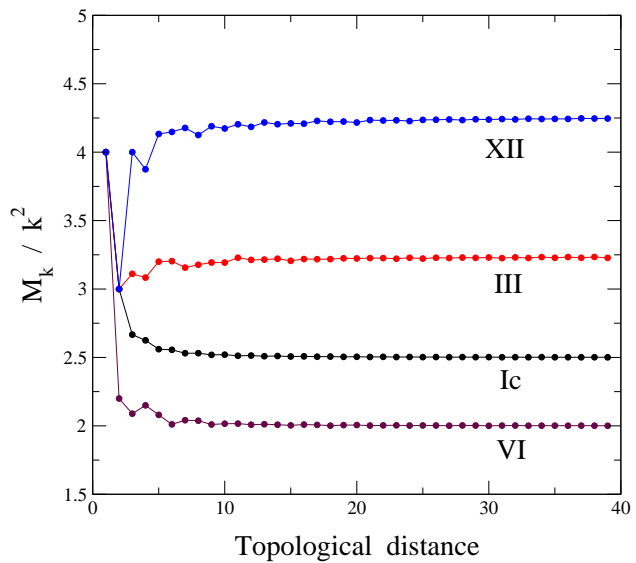}
\vspace{-0.7cm}
\caption{
Ratio $M_k / k^2$ vs topological distance $k$ for several ice
structures. For large $k$ this ratio converges to the coefficient
$a$ of the second-order term in Eq.~(\ref{nk}).
From top to bottom: ice XII, III, Ic, and VI.
}
\label{f2}
\end{figure}

As indicated above, for a parabolic dependence of $M_k$ as a function of
the distance $k$, the ratio $M_k/k^2$ should converge for large $k$ to
the coefficient $a$. For the ice structures considered here, this 
convergence is rather fast, as shown in Fig.~2 for several cases.
After some fluctuations for small $k$, for $k \gtrsim 10$ that ratio
converges rather smoothly to its high-distance limit.
We observe again the clear differences between different ice structures.
In particular, ices XII and VI have the largest and smallest $a$ value,
respectively. 
Note that the parabolic dependence of $M_k$ is in general not strictly
parabolic, although an equation such as Eq.~(\ref{nk}) can fit very well
the actual coordination sequences, with relative errors converging fast
to zero as $k$ increases.  

\begin{figure}
\vspace{-1.1cm}
\hspace{-0.5cm}
\includegraphics[width= 9cm]{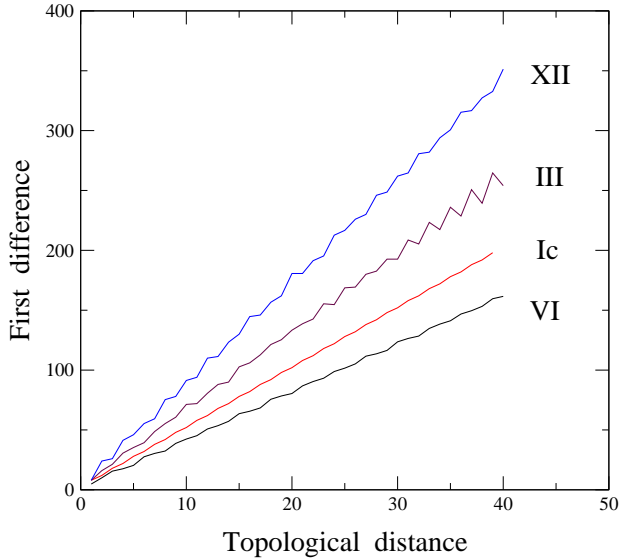}
\vspace{-0.7cm}
\caption{
First differences $\delta_k$ of the coordination sequence $\{M_k\}$
vs topological distance $k$ for several ice structures.
From top to bottom: ice XII, III, Ic, and VI.
}
\label{f3}
\end{figure}

To analyze with more detail the behavior of $M_k$ vs $k$ we 
consider the first and second differences, $\delta_k$ and $\epsilon_k$,
of the coordination sequences, defined in Eqs.~(\ref{dif1}) and
(\ref{dif2}), respectively.
For a strict parabolic dependence of the form given in Eq.~(\ref{nk})
one would expect a linear trend for the first differences:
 $\delta_k = 2 a k + a + b$.
The actual values of $\delta_k$ for various ice structures are shown 
in Fig.~3. We observe for ice Ic that $\delta_k$ follows closely a linear
dependence on $k$, with small fluctuations around a straight line of
slope $2 a$. Such fluctuations are larger for ice VI, and they become 
more prominent for ices III and XII. 

\begin{figure}
\vspace{-1.1cm}
\hspace{-0.5cm}
\includegraphics[width= 9cm]{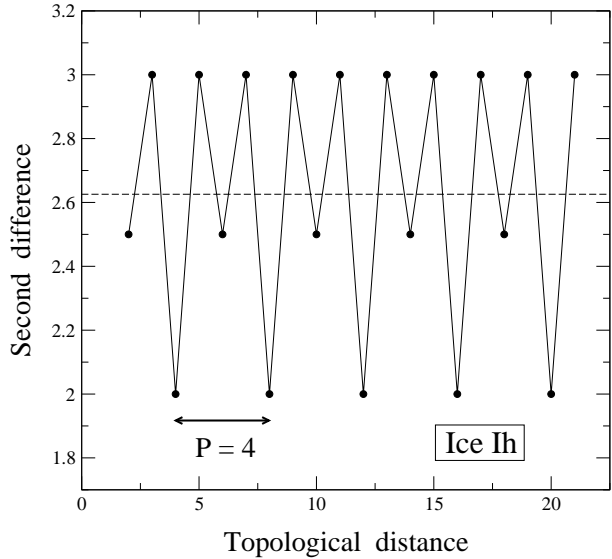}
\vspace{-0.7cm}
\caption{
Second differences $\epsilon_k$ of the coordination sequence vs
topological
distance for ice Ih.
The period length $P = 4$ is indicated.
The dashed line shows the parameter $a$ corresponding to this
structure.
}
\label{f4}
\end{figure}

We now go to the second differences of the sequence $\{ M_k \}$.
For a strict parabolic dependence, one would have a constant value for
the second differences: $\epsilon_k = a$.
In Fig.~4 we display $\epsilon_k$ as a function of $k$ for ice Ih.
The resulting $\epsilon_k$ fluctuates between 2 and 3, in a periodic 
sequence with 
period length $P = 4$. Note that the mean value of this sequence is
$a = 2.625$, which coincides with the value obtained from a direct fit 
for the sequence $M_k$.
The characteristics of the sequence $\{\epsilon_k\}$  depend strongly on 
the considered ice structure. Thus, in some cases one obtains a periodic
sequence as in the case of ice Ih, and in other cases one finds a
sequence $\{\epsilon_k\}$ for which no regular pattern is easily found.
The case of ice Ih is relatively simple in this respect, with a period
$P = 4$. Even simpler is the structure of ice Ic, for which $\{\epsilon_k\}$ 
alternates between two values (2 when $k$ is even and 3 when it is odd), 
with an average $a = 2.5$. Note that this structure has the same topology
as diamond, and is particularly simple to analyze. In fact, the exact
value of $M_k$ in this case is given by
\begin{eqnarray}
   M_k = \frac52 \, k^2 + \frac32 \, , \hspace{5mm} k = 2 n + 1  \\
   M_k = \frac52 \, k^2 + 2 \, , \hspace{5mm} k = 2 n + 2    \nonumber
\end{eqnarray}
for $n = 0, 1, 2, ...$

In other cases with a period length relatively small, 
it is not difficult to find exact expressions for $M_k$.
For example, for ice II, with period length $P = 4$, one has
(for $k > 5$):
\begin{equation}
   M_k = \frac72 \, k^2 - \frac12 \, k + c_k 
\end{equation}
where $c_k = 2$ for $k= 4 n$ ($n$ = 2, 3, 4, ...) and $c_k = 1$
otherwise.

\begin{figure}
\vspace{-1.1cm}
\hspace{-0.5cm}
\includegraphics[width= 9cm]{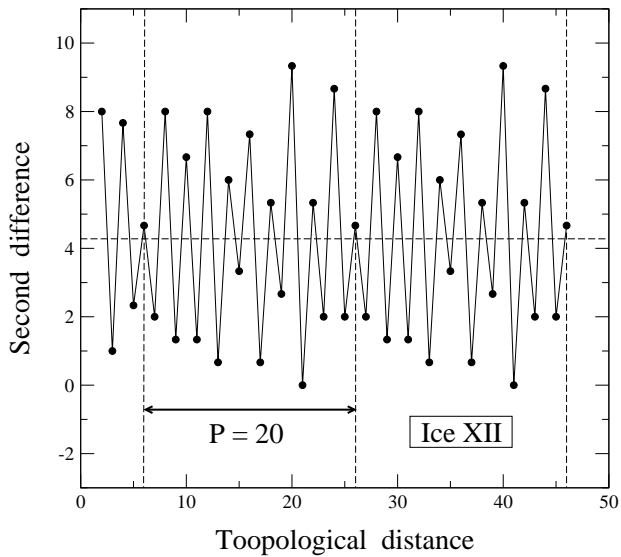}
\vspace{-0.7cm}
\caption{
Second differences $\epsilon_k$ of the coordination sequence
$\{M_k\}$ vs topological distance for ice XII.
The period length $P = 20$ is displayed.
The horizontal dashed line indicates the parameter $a$ corresponding
to this ice structure.
}
\label{f5}
\end{figure}

For other ice structures, we found sequences $\{\epsilon_k\}$
with longer period lengths. The longest of them corresponds to ice XII,
with $P = 20$. The sequence $\{\epsilon_k\}$ for this ice structure is
displayed in Fig.~5. One observes that a periodic sequence begins at
$k = 6$, and two complete periods appear in the figure until $k = 46$.
The mean value of $\epsilon_k$ over a period is 
$\bar{\epsilon}_k = 4.27$, in agreement with the $a$ value obtained 
from the parabolic fit of the sequence $\{ M_k \}$, and shown in
Table~III. 

As mentioned above, there are other ice structures for which we could
not find any repeated pattern in the sequence $\{\epsilon_k\}$.
As an example, we display in Fig.~6 $\{\epsilon_k\}$ for ice V.
We observe in this sequence that the values oscillate around the
corresponding parameter $a$ (horizontal dashed line), and the 
amplitude of the 
oscillations seems to increase as the distance $k$ becomes larger.
Among the ice structures considered here, we found three cases for 
which no periodic pattern appeared in the investigated $k$-region: 
ices III, IV, and V. This is indicated in the last column of Table~III 
as `NP'. For the other structures
(where a repeated pattern was found), a label `P' indicates
`periodic', followed by a number referring to the period length.
We note that two of the networks for which no periodic pattern 
in $\{\epsilon_k\}$ has been found, are in some sense special,
since ices III and V have been found to display partial 
hydrogen ordering. At present, we do not know any precise reason for this
behavior and, although not probable, we cannot exclude an accidental 
coincidence. This point should be investigated in the near future.

\begin{figure}
\vspace{-1.1cm}
\hspace{-0.5cm}
\includegraphics[width= 9cm]{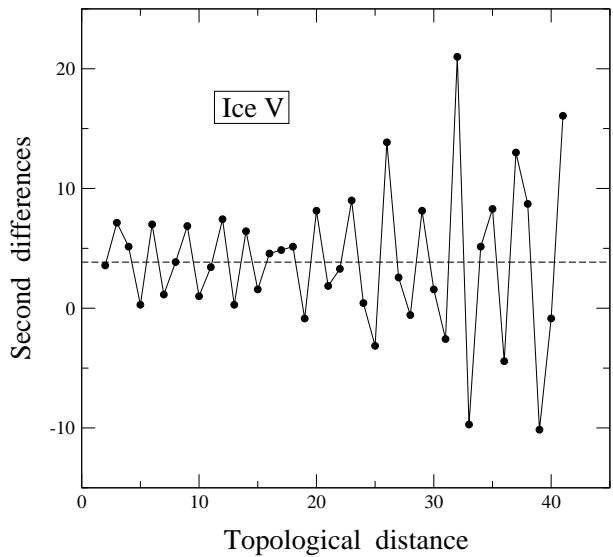}
\vspace{-0.7cm}
\caption{
Second differences $\epsilon_k$ of the coordination sequence vs
topological distance for ice V.
The dashed line indicates the parameter $a$ corresponding to this
structure.
}
\label{f6}
\end{figure}

Concerning the sequence $\{\epsilon_k\}$ , we emphasize that in the 
cases where no periodicity was found, the
existence of a period (maybe very long) cannot be excluded. In fact,
really long periods have been found in some zeolite structures, 
using an algebraic description.\cite{gr96}
However, this is not crucial for our present purposes of comparing
the topological density with other structural aspects of ice polymorphs.

\subsection{Topological density}

The long-distance behavior of the coordination sequence $\{M_k\}$ allows 
us to define the topological density of a given structure, as explained
in Sect.~II. The basic parameter for this purpose is the coefficient
$a$ of the quadratic term in Eq.~(\ref{nk}).
The topological density $\rho$ takes values ranging from 2.5 (ice Ic) to 
5 (ice VII) for the considered ice phases. In Table~III we give for each
structure the relative topological density $\rho / \rho_{\rm Ih}$ with 
respect to ice Ih.
Note that hexagonal ice Ih and cubic ice Ic, although very similar in
their local neighborhood have different topological density. Thus, for 
ice Ic $\rho / \rho_{\rm Ih} = 0.95$, in agreement with the lower $M_k$ 
values found for ice Ic, as compared to those of ice Ih.

 In connection with our results for the topological density $\rho$ of ice
polymorphs, it is worthwhile mentioning that  several authors have proposed 
to quantify the topological density in crystalline solids from the number 
of sites included in a cluster of a given topological radius,\cite{br93}
using expressions such as $S_k$ in Eq.~(\ref{sk}).
This definition, however, can lack an absolute meaning, as it yields 
values of the topological density which depend on the cluster size 
and on the normalization procedure.
In this line, a more precise definition for the topological density of 
crystalline solids was given earlier as\cite{gr96,eo04}
\begin{equation}
  \rho' = \lim_{k \to \infty} \frac{S_k}{k^D}
\end{equation}
($D$ = 3 here).
As indicated above, for large $k$ one has an asymptotic dependence for the 
mean coordination sequence: $M_k \sim a \, k^2$ (see Eq.~(\ref{mk})),
and therefore
\begin{equation}
   S_k = \sum_{i=1}^k M_i  \sim \frac13 \, a \, k^3
\end{equation}
as can be easily derived for the sum of squares of natural numbers.
In this way one has $\rho' = a / 3$.
Note that this definition, apart from a factor 3 in the denominator,
does not consider the number $w$ of disjoint subnetworks in a given
structure, as introduced in the present work.

Although one could in principle expect the existence of some relation
between topological density and molar volume of the different ice
structures, it is not evident that there is a close relation between
both quantities.
There are various reasons that can contribute to make difficult
this comparison. The most important is the dependence of the molar
volume on temperature and pressure in the parameter region where each
phase is stable (or metastable), whereas the topological density is
a fixed characteristic of each structure, irrespective of the value of
external thermodynamic or mechanical variables.
For this reason, we obtain a reference volume for each ice polymorph,
representative of the corresponding phase and independent of external
variables. Such a reference volume, $v_0$, can be considered as a 
fingerprint of the corresponding phase. 
To obtain $v_0$, we carry out a minimization procedure (at zero volume
and temperature), in which atomic positions and cell parameters are 
optimized using the q-TIP4P/F interaction model, as described in 
Sect.~II.

It can be argued that the use of an effective interatomic potential
may introduce a bias in the reference volume of the different ice
phases. This possible error, however, is not relevant for our present
purposes, as differences in cell volumes derived from the q-TIP4P/F
model and experimental data are smaller than 1\%, in particular at low
temperatures and relatively small external pressures.\cite{he11,ra12} 
A comparison of the volumes predicted by this and other interatomic
potentials, including density-functional theory calculations, has been
given elsewhere.\cite{pa12}
An advantage of the q-TIP4P/F model is that it allows us to optimize 
all degrees of freedom in relatively large unit cells with a reasonable 
computational effort. 

\begin{figure}
\vspace{-1.1cm}
\hspace{-0.5cm}
\includegraphics[width= 9cm]{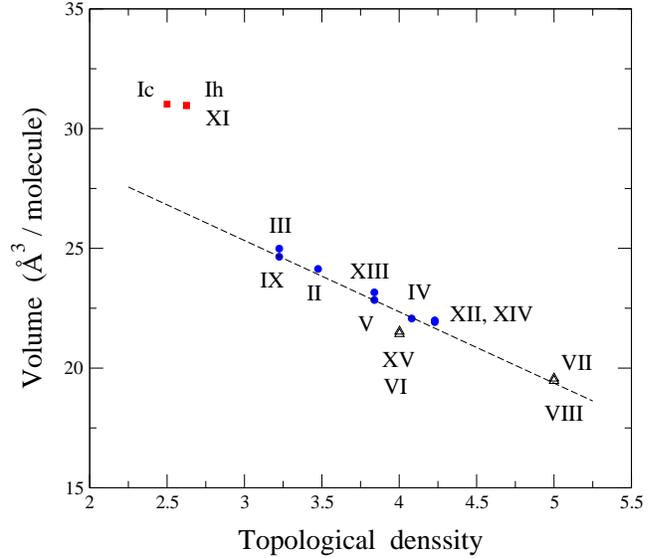}
\vspace{-0.7cm}
\caption{
Volume per molecule vs topological density $\rho$ for several ice
structures.
Data points corresponding to networks with $w = 2$ (ices VI, XV, VII,
and VIII) are represented by open triangles.
}
\label{f7}
\end{figure}

In Fig.~7 we present the volume per molecule $v_0$ vs the topological 
density $\rho$ for crystalline ice structures.
One observes that the data points are more or less aligned following a
straight line of negative slope. However, the data corresponding to ices
Ih, Ic, and XI depart from this line. This question will be discussed 
below.

It is important to recall that ice VII (along with ice VIII)
and ice VI (along with ice XV) contain two independent subnetworks.
This means that the total number of sites in each of these structures
is twice that of a single subnetwork. Since the parameter $a$ in these
cases takes into account only sites in one subnetwork, for these
structures the topological density $\rho$ equals $2 a$ instead of $a$
(see Sect.~II). Data points corresponding to cases with two subnetworks
($w = 2$) are displayed as open triangles in Fig.~7.
In connection with this, we remember that ice VII contains two
subnetworks, each of them equivalent to the ice Ic network, so that
the empty space has been filled by the two interpenetrating
networks and the volume per molecule correlates well with the
topological density, once the above-mentioned factor $w = 2$ is 
included.     In other words,
for each subnetwork of ice VII, the parameter $a$ coincides with that 
of ice Ic, but the topological density is twice larger in the former
case than in the latter, so that ice VII 
appears in Fig.~7 following the trend of other ice polymorphs in the 
$\rho - v_0$ plot. Something similar happens for ice VI.

According to the general trend observed in Fig.~7, one would expect 
for ices Ih and Ic (as well as ice XI) a volume per molecule smaller than 
that actually found.
This has to be related to the rigidity of the H-bridges in these
structures, along with the large empty space between water molecules,
which makes that ices Ih and Ic have fairly open low-density structures, 
where the packing efficiency is low.  Note that both ice structures
contain only six-membered rings. The difference between them
consists in the stacking sequence of layers formed by
water molecules arranged in six-membered rings\cite{ma09,sa11,pe99}
(`chair' or `boat' configurations).
In practice, structural differentiation between ices Ic and Ih is
complicated by the fact that ice Ic usually contains hexagonal stacking
faults.\cite{ar68,ko00,sa04b}

The stiffness of ice Ic and Ih structures can be related
to the rigidity of the tetrahedral arrangement of water molecules.
In this line, several orientational parameters have been defined
to measure the extent to which a molecule and its four nearest
neighbors adopt a tetrahedral arrangement.\cite{ch98,er01}
In particular, the order parameter $q$ employed in 
Refs.~\onlinecite{er01,mc12} gives for the ideal crystal structures
of ice Ih and Ic its maximum value $q = 1$, which means that 
tetrahedra are undistorted. 
For other crystalline ice structures, this parameter is clearly
lower than unity. 

Another crystalline ice structure containing open channels is that of
ice II, which presents one cavity per six water molecules.\cite{sa11}
The density of this ice polymorph is around 25\% higher than that of 
ice Ih, but the former structure contains more roomy cavities than
the latter.\cite{ma09} 
We note that the similarity between channels in ices Ih and
II is important for the mechanism of transition from one to the 
other.\cite{ma09}
However, concerning the rigidity of the tetrahedra in ice II, the
situation is clearly different than ice Ih, as in the former case 
the tetrahedral order parameter takes a value $q = 0.83$, clearly lower
than in the case of ice Ih. This indicates an appreciable distortion of
the tetrahedra of water molecules in the case of ice II.

In connection with the topological characterization of ice polymorphs,
an interesting question is its generalization to other phases, such
as amorphous ice and confined water.
For amorphous ice, in particular, a topological density $\rho$,
similar to that discussed here for crystalline phases, could be
defined. Given the three-dimensional character of the amorphous H-bond 
networks, one expects to find average coordination sequences $\{ M_k \}$ 
with a distance dependence similar to Eq.~(\ref{mk}).
This can allow to study the correlation between topological
density and actual molar volume of the amorphous phases.

\section{Concluding remarks}

We have presented an analysis of the ring statistics and coordination
sequences in crystalline ice structures. 
This has allowed us to quantitatively characterize the topology
of different ice polymorphs.
In particular, an analysis of the coordination sequences up to large
distances provides us with a quantitative measure of the topological 
density of the corresponding networks. 
This measure is obtained from the coefficient $a$ derived from the
parabolic dependence of the coordination sequence on the topological
distance.

We have found a correlation between the topological density $\rho$
and crystal volume of the considered ice polymorphs.
The general trend shown in Fig.~7 is apparently not followed by ices Ih 
and Ic, due to the low density and rigidity of these structures.

Other ways of characterizing the different ice networks can give
further insight into the structural properties of different polymorphs.
In this line, we mention the use of the so-called connective
constants, derived from self-avoiding walks in networks, to study
complex crystal structures such as those of zeolites.\cite{he95b}
All this can help to understand thermodynamic properties of solid water 
phases, as the configurational entropy associated to hydrogen disorder,
which is known to depend on the topology of ice
networks.\cite{he13,ma04b}  \\  \\

\begin{acknowledgments}
This work was supported by Direcci\'on General de Investigaci\'on 
(Spain) through Grant FIS2012-31713,
and by Comunidad Aut\'onoma de Madrid
through Program MODELICO-CM/S2009ESP-1691.
\end{acknowledgments}

\end {document}